\documentclass{iau} 
\usepackage{graphicx}
\usepackage{natbib}
\usepackage{aas_macros} 
\usepackage{gensymb}
\usepackage{float}

\title[Modeling Stellar Jitter for the Detection of Earth-Mass Exoplanets] 
{Modeling Stellar Jitter for the Detection\\of Earth-Mass Exoplanets via Precision Radial Velocity Measurements}

\author[Samuel Granovsky, et al.]   
{Samuel Granovsky\textsuperscript{1,2,3}, Irina N. Kitiashvili\textsuperscript{1}, Alan Wray\textsuperscript{1}}

\affiliation{{\textsuperscript{1}NASA Ames Research Center, Moffett Field, MS 258-6,
	Mountain View, CA, USA} \\ {\textsuperscript{2}New Jersey Institute of Technology, 323 Dr Martin Luther King Jr Blvd, Newark, NJ, USA} \\ {\textsuperscript{3}Universities Space Research Association, 7178 Columbia Gateway Drive, Columbia, MD, USA}}	
	
\pubyear{2022}
\volume{362}  
\jname{Predictive Power of Computational Astrophysics\\as a Discovery Tool}
\editors{D.V. Bisikalo, ed.}

\begin{document}
\maketitle

\begin{abstract}
The detection of Earth-size exoplanets is a technological and data analysis challenge. Future progress in Earth-mass exoplanet detection is expected from the development of extreme precision radial velocity measurements. Increasing radial velocity precision requires developing a new physics-based data analysis methodology to discriminate planetary signals from host-star-related effects, taking stellar variability and instrumental uncertainties into account. In this work, we investigate and quantify stellar disturbances of the planet-hosting solar-type star HD121504 from 3D radiative modeling obtained with the StellarBox code. The model has been used for determining statistical properties of the turbulent plasma and obtaining synthetic spectroscopic observations for several Fe I lines at different locations on the stellar disk to mimic high-resolution spectroscopic observations. 
\end{abstract}

\begin{figure}[b]
	\begin{center}
		\includegraphics[width=5.3in]{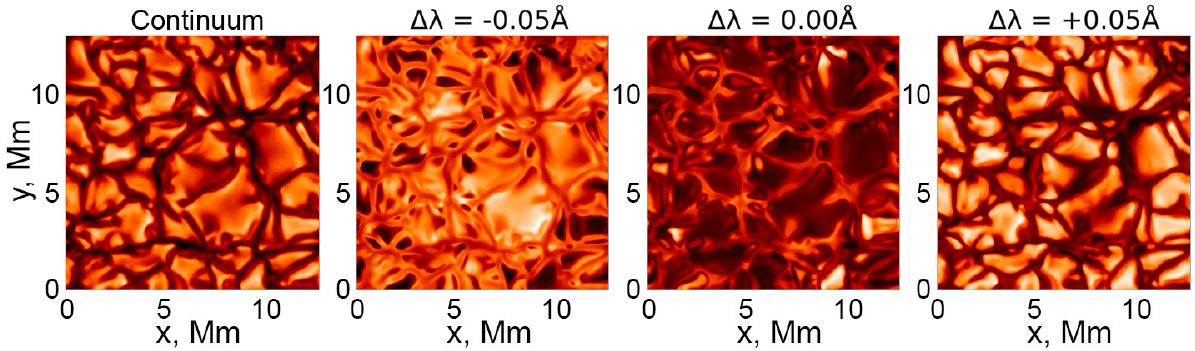} 
	\end{center}
	\caption{Intensity variations at different locations from the Fe I line core ($\lambda$\textsubscript{ref} = 6173.7\AA) show how the structure of the atmosphere changes with height. The presented synthetic images correspond to an area at the disk center.}
	\label{fig:intensity}
\end{figure}

\section{Introduction}
Due to the relatively large mass of Jupiter-size planets, the radial velocity (RV) method for detecting exoplanets is viable since such planets are able to apply a sufficient acceleration to their host star. This is especially true for Jupiter-mass planets which orbit close to their host stars. However, this is typically not the case for smaller Earth-mass exoplanets, especially those which orbit farther from their host stars. Variations in RV of the host star as it orbits the barycenter is comparable to the fluctuations in RV caused by stellar jitter, the noise caused by the movement of material on the star's surface. While noise caused by stellar jitter may be on the order of 100 m/s \citep{Saar1997} in relation to the star's center of mass, the RV of the star due to the effects of an Earth-mass planet may be only 0.1 to 1 m/s \citep{Plavchan2015}. Therefore, development of an accurate model of stellar jitter is essential for making RV measurements precise enough to detect Earth-mass exoplanets. This paper presents initial modeling results of the stellar jitter for star HD121504 using 3D radiative simulations and a radiative transfer code to generate time series for synthetic observables.   

\begin{figure}[t]
	\begin{center}
		\includegraphics[width=4.5in]{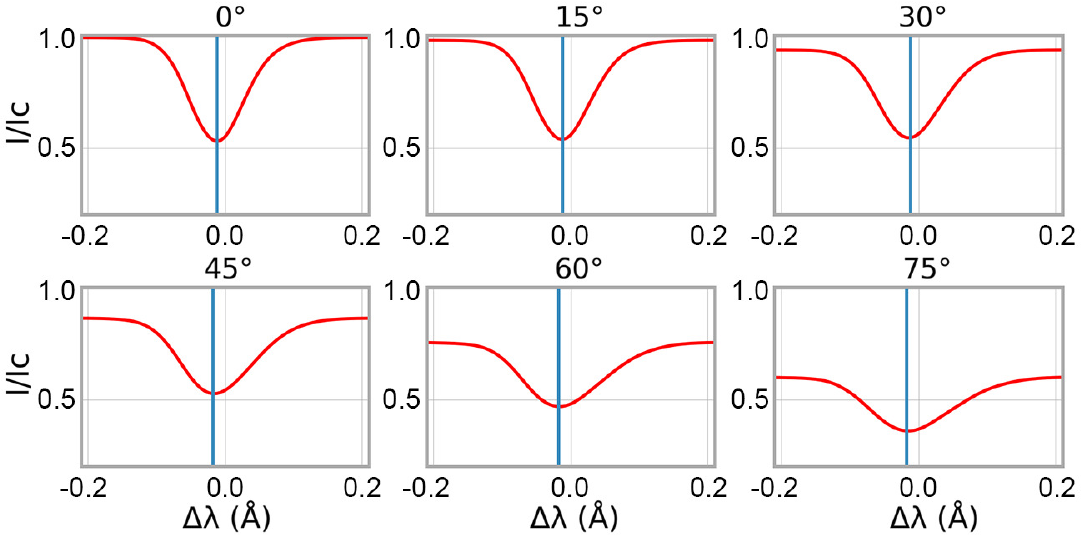} 
		\caption{Spectral line profiles for the Fe I line ($\lambda$\textsubscript{ref} = 6173.7\AA, red) at several angular distances from disk center. Vertical lines indicate the location of the line core.}
		\label{fig:line_profiles}
	\end{center}
\end{figure}

\begin{figure}[b]
	\begin{center}
		\includegraphics[width=5in]{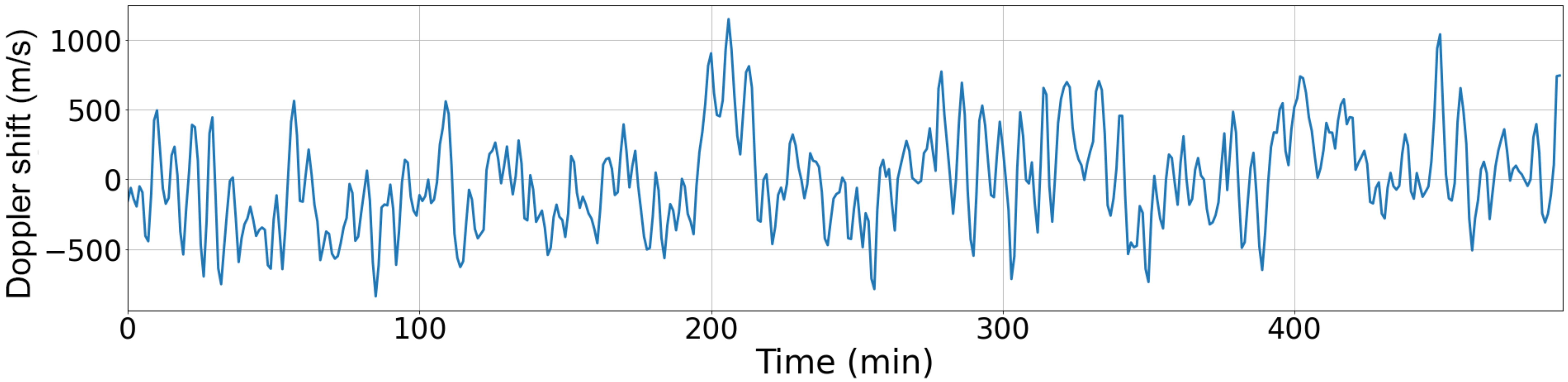} 
		\caption{Disk-averaged Doppler-shift time variations obtained for star HD12504.}
		\label{fig:doppler}
	\end{center}
\end{figure}

\section{Modeling of Synthetic Observables}
To investigate convection-driven disturbances, we obtained a 3D radiative model of star HD121504 via the StellarBox code \citep{Wray2015,Wray2018}. The stellar surface area 12.8 Mm x 12.8 Mm wide with a spatial resolution of 50 km were computed centered on the following Fe I lines: 5247Å, 5250.6Å, 5251Å, 6173.7Å, 6301.9Å, and 6302.9Å. The spectral line synthesis has been performed using the SPINOR code \citep{Frutiger2000} for every 10\degree{} in angular distance between -80\degree{} and +80\degree{} from the disk center and every 30\degree{} in polar angle between -60\degree{} and +90\degree{}. An example of a snapshot at disk center is shown in Figure~\ref{fig:intensity} at various wavelengths relative to the reference line at 6173.7Å. Figure~\ref{fig:line_profiles} illustrates how properties of a spectral line change between disk center and limb. In particular, there is a decrease of the continuum intensity and an increase of the full width at half maximum (FWHM) as flows tangent to the stellar surface gain a radial component relative to the observer. 

Using the described process, a time-series comprising 492 minutes of stellar dynamics was computed for the six Fe I lines for each of the previously mentioned locations. For each line, a weighted disk-averaged profile was computed. From the center of mass of each profile, Doppler-shift time-series were obtained. Figure~\ref{fig:doppler} shows an example of the disk-averaged Doppler shift derived from synthetic data of 6173.7\AA. Preliminary results show Doppler shift fluctuations for HD121504 mostly in the range of $\pm 500$~m/s.

\begin{figure}[t]
	\begin{center}
		\includegraphics[width=4.5in]{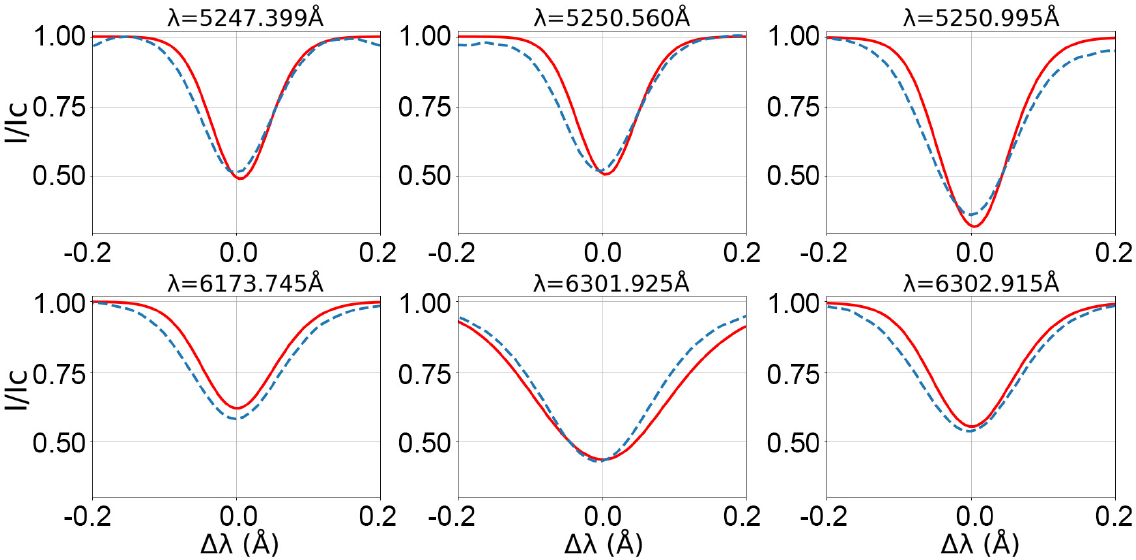} 
		\caption{Comparison of observed (blue, dotted) and synthetic (red, solid) spectral lines of Fe I.}
		\label{fig:comparison}
	\end{center}
\end{figure}

The disk-averaged profiles were compared to observational data of HD121504 taken by the ESO HARPS spectrograph for each of the six lines, as shown in Figure~\ref{fig:comparison}. The observed profiles (blue curves) differ somewhat from the synthetic profiles (red), which is potentially due to low spacial resolution of the numerical model, as well as magnetic effects not yet being accounted for. There is also some uncertainty in the absolute iron abundance that should be used to perform line synthesis.

\section{Discussion and Conclusions}
Detection of Earth-mass exoplanets orbiting solar-type stars is challenging due to significant contamination of the RV signal with disturbances originating from turbulent photospheric dynamics. To model these disturbances, we use a 3D radiative model of the planet-hosting star HD121504 to generate synthetic observables and to characterize stellar convective motions. To characterize the stellar convective motions near the photosphere, we computed synthetic high-resolution spectra of six Fe I lines for different locations on the stellar disk to obtain disk-integrated observables . Further analysis using hydrodynamic and MHD simulations and synthetic data sets with higher spatial and temporal resolutions is required for a more realistic representation of observational data. For future disk-averaged profiles, we plan to apply a randomized temporal offset to the time-series for each location on the disk to prevent resonance affecting the Doppler data, as the current model uses a disk average which assumes every location on the disk behaves identically at any given time.

{\bf Acknowledgments.} Observations used in the paper were made with HARPS spectrograph on the 3.6m ESO telescope at La Silla Observatory, Chile. This work is supported by the NASA Extreme Precision Radial Velocity Foundation Science Program.


\end{document}